\documentclass[12pt]{article}
\usepackage{cite}
\usepackage{amsmath,amssymb,amsfonts,mathtools}
\usepackage{algorithmic}
\usepackage{graphicx}
\usepackage{textcomp}
\usepackage{xcolor}
\usepackage{amsthm}
\usepackage{mathabx}
\usepackage{hyperref}
\usepackage{caption}
\usepackage{tablefootnote}
\usepackage{array}
\usepackage{mathrsfs}

\usepackage{xpatch}
\makeatletter
\AtBeginDocument{\xpatchcmd{\@thm}{\thm@headpunct{.}}{\thm@headpunct{}}{}{}}
\makeatother

\DeclareCaptionType{equ}[][]
\newtheorem{theorem}{Theorem}[section]

\newtheorem{prop}[theorem]{Proposition}
\newtheorem{cor}[theorem]{Corollary}
\newtheorem{example}[theorem]{Example}
\newtheorem{remark}[theorem]{Remark}

\begin{document}

\title{The Dimensions of the Hulls of Conorm Codes from Algebraic Geometry Codes}
	
\author{Junmin An\\Department of Mathematics\\Sogang University, Seoul, Korea\\ {\tt junmin120@sognag.ac.kr}\\
\\Jon-Lark Kim \\ Department of Mathematics \\ Sogang University, Seoul, Korea \\
		{\tt jlkim@sogang.ac.kr } \\
	}
\date{November 2, 2023}
	
\maketitle

\begin{abstract}
Chara et al. introduced conorm codes defined over algebraic geometry codes, but the hulls of conorm codes were not determined yet. In this paper, we study the dimension of the hull of conorm codes using the method introduced by Camps et al. For an algebraic geometry code $\mathcal{C}:=C_\mathscr{L}(D, G)$, we consider the divisor $\gcd(G, H)$, where $H$ is the divisor satisfying
\[C_\mathscr{L}(D, G)^\perp=C_\mathscr{L}(D, H).\]
Given an extension $F'/\mathbb{F}_{q^t}$ of an algebraic function field $F/\mathbb{F}_q$, we assume that the divisor $\gcd(G, H)$ is non-special. If the degree of $\gcd(G, H)$ is greater than $2g-2+{t\over [F':F]}\deg\text{Diff}(F'/F)$, then we have determined the exact dimension of the hull of the conorm of $\mathcal{C}$. If not, we have determined the lower bound of the dimension of the hull of the conorm of $\mathcal{C}$. We provide some examples for the dimension of the hull of certain conorm codes of AG codes defined over a rational function field.

\end{abstract}

\section{Introduction}

Algebraic geometric codes were first introduced by Goppa using algebraic curves over finite fields~\cite{Goppa}. As algebraic geometry codes are a generalization of Reed-Solomon codes, AG code inherit the characteristic of enabling the construction of a code whose minimum distance exceeds the designed distance.  Furthermore, it has been observed that algebraic geometry codes can surpass the Gilbert-Varshamov bound~\cite{Tsfasman}, and this aroused lots of interests in the field of algebraic geometry codes.

Assmus and Key~\cite{Assmus} introduced the concept of the hull of a linear code, defined as the intersection of a code and its dual. The hull determines the complexity of the algorithm for checking permutation equivalence of two linear codes~\cite{Kim},\cite{Qian} or computing the automorphism group of linear codes~\cite{Qian}. The complexity of these problems plays an important role in cryptography. Notably, Massey~\cite{Massey} introduced the notion of LCD codes, which are linear codes characterized by an empty hull. S. Mesnager~\cite{Mesnager} proposed the method for constructing LCD codes from algebraic curves. Camps et el.~\cite{Camps} extended Mesnager's idea to determine the hulls of algebraic geometry codes.

Chara et al.~\cite{Chara} devised conorm codes from algebraic geometry codes using the conorm map, which is a homomorphism from the divisor group of a function field  $F$ to the divisor group of a finite algebraic extension of $F$. They studied the parameters, specifically lengths dimensions, and minimum distances of conorm codes.

The hull of a linear code is an important property which characterizing the code, but the hulls of conorm codes have not been studied yet. So, we have studied the hull of conorm codes in our paper. We have used the method introduced by Camps et el.~\cite{Camps} to determine the dimension of the hull of conorm codes. Given an algebraic geometry code $\mathcal{C}=\mathcal{C}_\mathscr{L}(D, G)$ associated with divisors $D$ and $G$ of a function field $F$ whose dual is $\mathcal{C}_\mathscr{L}(D, H)$, if $\gcd(G, H)$ is non-special, then we show how the dimension of the hull of the conorm of $\mathcal{C}$, defined over $F'$, is related to the dimension of the hull of $\mathcal{C}$. Specifically, if the degree of $\gcd(G, H)$ is greater than $2g-2+{t\over m}\deg\text{Diff}(F'/F)$, where $\text{Diff}(F'/F)$ is the different divisor of $F'$ over $F$, then we have obtained the exact hull dimension of the conorm of $\mathcal{C}$. Otherwise, we have obtained the lower bound of the hull dimension of the conorm of $\mathcal{C}$. Since any univariate function field is a finite algebraic extension of a rational function field, we can represent some AG codes as conorm codes of rational AG codes. We present some examples of the hull dimension of conorm codes of rational AG codes for some well known function fields, namely elliptic, hyperelliptic, and hermitian function fields.

Our paper is organized as follows. In Section 2, we review some concepts of algebraic function fields and algebraic geometry codes. In Section 3 and 4, we present our results regarding the dimension of the hull of conorm codes for unramified extensions and separable ramified extensions, respectively. In Section 5 we provide some examples of the dimension of the hull of conorm codes of rational AG codes, and we conclude our paper in Section 6.

\section{Preliminaries}

Here we review some concepts of algebraic function fields and algebraic geometry codes from~\cite{Stichtenoth}.

Let $F$ be a univariate function field over a finite field $\mathbb{F}_q$. The divisor group of $F$ is a free abelian group generated by the places of $F$. Elements of the divisor group are represented as $D=\sum\limits_{P}n_P\cdot P$. For a given divisor $D$, we define the support of $D$ by $\text{supp}D:=\{P|\alpha_P\ne 0\}$ and the degree of $D$ by $\deg D:=\sum\limits_Pn_P\cdot\deg P$.

The {\em Riemann-Roch space} of a divisor $D$ is defined by
\[\mathscr{L}(D):=\{x\in F|(x) \ge -D\}\]
where $(x)$ is the principal divisor of $x$. As $\mathscr{L}(D)$ is the vector space over $\mathbb{F}_q$, the Riemann-Roch Theorem states that the dimension of $\mathscr{L}(D)$ as a $\mathbb{F}_q-$vector space is given by
\begin{equation}
\dim \mathscr{L}(D)=\deg D+1-g+\dim\mathscr{L}(W-D)
\end{equation}
where $W$ is any canonical divisor of $F$ and $g$ is the genus of $F$. We denote the dimension of $\mathscr{L}(D)$ by $\ell(D)$.

The index of specialty of the divisor $D$ is given by $i(D):=\ell(D)-\deg D+g-1$. By the Riemann-Roch Theorem, we can easily determine that $i(A)=0$ if and only if $\ell(W-D)=0$ for any canonical divisor $W$ of $F$. A divisor $D$ with $i(D)=0$ is called a non-special divisor; otherwise, it is referred to as a special divisor. The Following are some properties of the index of specialty:
\begin{enumerate}
\item[(1)] If $\deg D > 2g-2$, then 	$D$ is a non-special divisor.
\item[(2)] If $\deg D < g-1$, then $D$ is a special divisor.
\end{enumerate}

Let $D$ be a given divisor. If $\deg D> 2g-2$, then we have
\[\ell(D)=\deg D+1-g\]
by the property of the index of specialty and the Riemann-Roch Theorem. On the other hand, if $\deg D<0$, then $\ell(D)=0$. In the case where $0 \le \deg D \le 2g-2$, the dimension of the Riemann-Roch space of $D$ does not depend solely on the degree of $D$, but still, we can make some observations about $\ell(D)$ with respect to $\deg D$.
\begin{theorem}~\cite{Stichtenoth}
For a divisor $D$ with $0 \le \deg D\le 2g-2$,
\[2(\ell(D)-1)\le \deg D.\]
\end{theorem}
By the Clifford's Theorem, if $\deg D=0$, then $\ell(D)$ is either $0$ or $1$. Indeed, if $\deg D=0$, then $\ell(D)=0$ if and only if $D$ is a principal divisor.

Now, we consider the conorm map on the divisor group. Let $F'/\mathbb{F}_{q^t}$ be an algebraic extension of $F/\mathbb{F}_q$. For a place $P$ of a function field $F$, the conorm of $P$ is defined by
\[\text{Con}_{F'/F}(P):=\sum\limits_{P'|P}e(P'|P)\cdot P'\]
where $e(P'|P)$ is the ramification index of $P'$ over $P$. This definition of the conorm map naturally extends to a homomorphism from the divisor group of $F$ to the divisor group of $F'$. For a divisor $D=\sum\limits_{P}n_P\cdot P$ of $F$, the conorm of $D$ is defined by
\[\text{Con}_{F'/F}(D):=\sum\limits_P\sum\limits_{P'|P}n_P\cdot e(P'|P)\cdot P'.\]
The conorm map on the divisor group preserves the principal divisors. For a nonzero $x\in F$, let $(x)^F$ be the principal divisor of $x$ in $F$ and $(x)^{F'}$ be the principal divisor of $x$ in $F'$. Then, we have
\[\text{Con}_{F'/F}((x)^F)=(x)^{F'}.\]
The degree of the conorm divisor is given by the immediate consequence of the following theorem.
\begin{theorem}~\cite{Stichtenoth}
	Let $F'/\mathbb{F}_{q^t}$ be a finite algebraic extension of $F/\mathbb{F}_q$. For a place $P$ of $F$ and a place $P'$ of $F'$ which is lying over $P$, let $e(P'|P)$ and $f(P'|P)$ be the ramification index and the relative degree of $P'$ over $P$ respectively. Then
	\[\sum\limits_{P'|P}e(P'|P)\cdot f(P'|P)=[F':F].\]
\end{theorem}
By above equation, we can verify that for a divisor $D$ of $F$,
\[\deg \text{Con}_{F'/F}(D)={[F':F]\over t}\deg D.\]

For an algebraic extension $F'/\mathbb{F}_{q^t}$ of $F/\mathbb{F}_q$, the different divisor of $F'$ over $F$ is defined by
\[\text{Diff}(F'/F):=\sum\limits_P\sum\limits_{P'|P}d(P'|P)\cdot P'\]
where $d(P'|P)$ is the different exponent of $P'$ over $P$.

\begin{theorem}~\cite{Stichtenoth}
	Let $F'/\mathbb{F}_{q^t}$ be an algebraic extension of algebraic function field $F/\mathbb{F}_q$. For a place $P'$ of $F'$ lying over a place $P$ of $F$, 
\[d(P'|P)\ge e(P'|P)-1\]	
and the equality holds if and only if $e(P'|P)$ is not divisible by the characteristic $p$.
\end{theorem}

The genus of $F'$ is determined by the following theorem.
\begin{theorem}~\cite{Stichtenoth}
	Let $F/\mathbb{F}_q$ be an algebraic function field with genus $g$ and $F'/\mathbb{F}_{q^t}$ be a finite separable extension of $F$. Let $g'$ be the genus of $F'$. Then, the following holds.
	\[2g'-2={[F':F]\over t}(2g-2)+\deg \textup{Diff}(F'/F).\]
\end{theorem}
In case that $F'$ is an unramified separable extension of $F$, the degree of the different divisor of $F'$ over $F$ is zero.

Let $P_1,\ldots P_n$ be rational places of an algebraic function field $F$. Let $D=P_1+\cdots+P_n$ and $G$ be a divisor of $F$ with support disjoint from $D$. Then, the algebraic geometry code $C_\mathscr{L}(D, G)$  is defined as
\[C_\mathscr{L}(D, G):=\{x(P_1),\ldots,x(P_n)|x\in \mathscr{L}(G)\}.\]
For an algebraic geometry code $C_\mathscr{L}(D, G)$, there always exists a divisor $H$ such that $C_\mathscr{L}(D, H)$ is the dual of $C_\mathscr{L}(D, G)$.

The greatest common divisor of two divisors $D_1:=\sum\limits_Pn_P\cdot P$ and $D_2=\sum\limits_Pm_P\cdot P$ is defined as
\[\gcd(D_1, D_2):=\sum\limits_P\min(n_P, m_P)\cdot P\]
By the definition of the hull of a linear code,
\[C_\mathscr{L}(D, \gcd(G, H))\subseteq \textup{Hull}(C_\mathscr{L}(D, G)).\]
The Following proposition states that equality holds under certain conditions.

\begin{prop}~\cite{Camps}
For a given algebraic geometry code $C_\mathscr{L}(D, G)$ defined over an algebraic function field $F$, let H be the divisor of $F$ such that
\[C_\mathscr{L}(D, G)^\perp=C_\mathscr{L}(D, H).\]
If $2g-2<\deg G<n$ and $\gcd(G, H)$ is a non-special divisor, then
\begin{equation}
	\textup{Hull}(C_\mathscr{L}(D, G))=C_\mathscr{L}(D, \gcd(G, H)).
\end{equation}

\end{prop}

Now we define the conorm of an AG code. Let $F'/\mathbb{F}_{q^t}$ be an algebraic extension of an algebraic function field $F/\mathbb{F}_q$ of degree $m$. For each place $P\in\text{supp}D$ of F, let $m_P$ be the number of places of $F'$ lying over $P$. We assume that
\begin{equation}
e(P'|P)={m\over m_P}	
\end{equation}
holds for each place $P'$ of $F'$ lying over $P$. Given an algebraic geometry code $\mathcal{C}:=C_\mathscr{L}(D, G)$ defined over $F$, ~\cite{Chara} defines the conorm of $\mathcal{C}$ by
\[\text{Con}_{F'/F}(\mathcal{C}):=C_\mathscr{L}(D', G')\]
where
\begin{equation}
	D'={1 \over m}\sum\limits_{P\in\text{supp}(D)}m_P\cdot\text{Con}_{F'/F}(P)\quad\text{and}\quad G'=\text{Con}_{F'/F}(G).
\end{equation}
Clearly the two divisors $D'$ and $G'$ are distinct.
If $F'/F$ is an unramified extension of degree $m$ with $t=1$, and $(q, m)=1$, then~\cite{Chara} showed that the following holds.
\begin{equation}
	\text{Con}_{F'/F}(C^\perp)=\text{Con}_{F'/F}(C)^\perp.
\end{equation}
This may holds even if the given conditions are not satisfied. In the case that the extension $F'/F$ is separable, not necessarily unramified, (5) holds only if
	\[\displaystyle{1\over t}\left(mn-\sum\limits_{P\in\text{supp} D}m_P\right)=\deg\text{Diff}(F'/F).\]

\section{The hull of conorm codes on unramified extension}
In this section, we show how the dimension of the hull, or simply the hull dimension, of conorm codes is determined in the case that the extension $F'/\mathbb{F}_{q^t}$ over $F/\mathbb{F}_q$ is unramified and $t=1$, i.e., $\mathbb{F}_{q^t}=\mathbb{F}_q$. When considering the hulls of linear codes, LCD codes and self-dual codes are two extreme cases. That is, the hull of an LCD code is a zero vector space and the hull of a self-dual code is identical to the code itself. We consider these two cases separately. We use the notation $h(\mathcal{C})$ to represent the hull dimension of a linear code $\mathcal{C}$.
\begin{prop}
Let $F'$ be an unramified finite extension of an algebraic function field $F/\mathbb{F}_q$ of degree $m$ with $(q,\,m)=1$. Let $\mathcal{C}$ be an algebraic geometry code defined on $F$. If $\mathcal{C}$ is self-dual, then $\textup{Con}_{F'/F}(\mathcal{C})$ is also self-dual.
\end{prop}
\proof It immediately follows from (5).

\begin{theorem}
Let $F'$ be an unramified finite separable extension of an algebraic function field $F/\mathbb{F}_q$ of degree $m$ with $(q,\,m)=1$, and let $\mathcal{C}:=C_\mathscr{L}(D, G)$ be an algebraic geometry code defined over $F$ with $2g-2<\deg G<n$. Assume that $H$ is the divisor of $F$ such that 
\[C_\mathscr{L}(D, G)^\perp=C_\mathscr{L}(D, H).\]
 Let the divisor $\gcd(G, H)$ is non-special. Then the dimension of the hull of $\mathcal{C'}:=\textup{Con}_{F'/F}(\mathcal{C})$ satisfies
 	\[h(\mathcal{C'})\ge m\cdot h(\mathcal{C}),\]
and if $2g-2<\deg(\gcd(G, H))$, then
 	\[h(\mathcal{C'})= m\cdot h(\mathcal{C}).\]
\end{theorem}

\proof Let $G'$ and $H'$ be the conorm of $G$ and $H$ respectively. Note that \[\gcd(G',\, H')=\sum\limits_P\min(\nu_P(G),\,\nu_P(H))\cdot \text{Con}_{F'/F}(P)=\gcd(G,\,H)'\]
where $\gcd(G,\,H)'$ is the conorm of $\gcd(G,\,H)$.

By (1), 
\[\ell(\gcd(G', H'))=\ell(\gcd(G, H)')={m}\cdot \ell(\gcd(G, H))+\ell(W-\gcd(G, H)'),\]
where $W$ is any canonical divisor of $F'$. So,
\begin{align*}
	h(\mathcal{C'}) &\ge \ell(\gcd(G, H)')\\&\ge{m}\cdot \ell(\gcd(G, H))\\&={m}\cdot h(\mathcal{C}).
\end{align*}
If $2g-2<\deg(\gcd(G, H))$, then,
\[2g'-2={m}(2g-2)<{m}\cdot \deg(\gcd(G, H))=\deg(\gcd(G, H)').\]
where $g'$ is the genus of $F'$. This implies that $\gcd(G, H)'$ is a non-special divisor, and so, 
\[\ell(\gcd(G, H)')={m}\cdot \ell(\gcd(G, H)).\]
Hence,
\[h(\mathcal{C'})={m}\cdot h(\mathcal{C}).\]

\begin{theorem}
	Let $F'$ be an unramified finite extension of an elliptic function field $F/\mathbb{F}_q$ of degree $m$ with $(q,\,m)=1$, and let $\mathcal{C}:=C_\mathscr{L}(D, G)$ be an algebraic geometry code defined over $F$ with $0<\deg G<n$. Assume that $H$ is the divisor of $F$ such that 
\[C_\mathscr{L}(D, G)^\perp=C_\mathscr{L}(D, H).\]
 Let $\gcd(G, H)$ be a non-special divisor. If $\mathcal{C}$ is an LCD code, then $\textup{Con}_{F'/F}(\mathcal{C})$ is also an LCD code.
\end{theorem}

\proof If $\gcd(G, H)'$ is non-special, then it is trivial by the previous theorem. So we assume that $\gcd(G, H)'$ is special. By Theorem 1,
\[\ell(\gcd(G, H)')\le 1+{1 \over 2}\cdot \deg(\gcd(G, H))=1.\]
Since $\gcd(G, H)'$ is special, $\ell(\gcd(G, H)')$ must be $1$. This implies that $\gcd(G, H)'$ is a principal divisor, as stated in Corollary 1.4.12. of \cite{Stichtenoth}. By the property of the conorm map that preserves principal divisors, $\gcd(G, H)$ is also a principal divisor. Since this is a contradiction, $\gcd(G, H)'$ must be a non-special divisor.

\begin{remark}
	For an algebraic function field $F/\mathbb{F}_q$, let $F'=F\mathbb{F}_{q^t}$ with $(q, t)=1$. Let $\mathcal{C}$ be an algebraic geometry code defined over $F$. Then
	\[h(\textup{Con}_{F'/F}(\mathcal{C}))=h(\mathcal{C}).\]
Indeed, the generator matrix $G'$ of $\textup{Con}_{F'/F}(\mathcal{C}))$ is given by
\[G'=G\otimes [\underbrace{1 , 1 , \dotsc, 1}_{t}]\]
\end{remark}

\proof Since $F'$ is a constant extension of $F$, $F'/F$ is an unramified extension such that $\ell(\text{Con}_{F'/F}(D))=\ell(D)$ and $\deg \text{Con}_{F'/F}(D)=\deg D$ for any divisor $D\in\text{Div}(F)$.

Note that any basis of $\mathscr{L}(G)$ is contained in some basis of $\mathscr{L}(\text{Con}_{F'/F}(G))$. This implies that any basis of $\mathscr{L}(G)$ is also basis of $\mathscr{L}(\text{Con}_{F'/F}(G))$. Given a place $P$ of $F$, let $P_i$ be any extension of $P$. Then,
\[v_P(x)=v_{P_i}(x)\]
for all $x\in F$ since $F'/F$ is unramified. So we conclude that
	\[G'=G\otimes [\underbrace{1 , 1 , \dotsc, 1}_{t}].\]
Then, by~\cite{Li}, the hull dimension of $\mathcal{C}'$ is given by
\begin{align*}
h(\mathcal{C}')&=\dim(\mathcal{C}')-\text{rank}(G'(G')^T)\\&=\dim(\mathcal{C})-\text{rank}(t\cdot GG^T)\\&=h(\mathcal{C}).
\end{align*}

\section{The hull of conorm codes on ramified extension}
In this section, we assume that the extension $F/\mathbb{F}_{q^t}$ over an algebraic function field $F/\mathbb{F}_q$ is separable.

We say that the extension $F/\mathbb{F}_{q^t}$ over a function field $F/\mathbb{F}_q$ is ramified if there is a place $P$ of $F$ which is ramified in $F'/F$. There are two types of a ramified extension: we say that a ramified place $P$ is tamely ramified if the characteristic of $\mathbb{F}_q$ does not divide the ramification index of any extension of $P$. Otherwise, we say that the place $P$ is wildly ramified. If there is no wildly ramified place, then we say that $F'$ is a tamely ramified extension of $F$.

If $F'/F$ is finite, separable extension, then the number of ramified places is finite.

\begin{theorem}
Let $F'/\mathbb{F}_{q^t}$ be a finite separable extension of an algebraic function field $F/\mathbb{F}_q$ of degree $m$. For an algebraic geometry code $\mathcal{C}:=C_\mathscr{L}(D, G)$ defined over $F$ with $2g-2<\deg G<n$, assume that $H$ is the divisor of $F$ such that 
\[C_\mathscr{L}(D, G)^\perp=C_\mathscr{L}(D, H).\]
 Let the divisor $\gcd(G, H)$ be non-special. If the extension $F'/F$ and the code $\mathcal{C}$ satisfies (3) and (5), then the hull of $\mathcal{C'}:=\textup{Con}_{F'/F}(\mathcal{C})$ satisfies,
		\[h(\mathcal{C'})\ge{m \over t}\cdot h(\mathcal{C})-{1\over 2}\deg\textup{Diff}(F'/F),\]
and if
	\[\deg(\gcd(G, H))>2g-2+{t\over m}\deg\textup{Diff}(F'/F),\]
then,
		\[h(\mathcal{C'})={m \over t}\cdot h(\mathcal{C})-{1\over 2}\deg\textup{Diff}(F'/F).\]
\end{theorem}

\proof By Theorem 4, the genus $g'$ of $F'$ is given by
\[2g'-2={m \over t}(2g-2)+\deg\textup{Diff}(F'/F).\]
So, the hull dimension of $\mathcal{C'}$ satisfies
\begin{align*}
	h(\mathcal{C'}) &\ge \ell(\gcd(G, H)')\\&\ge \deg(\gcd(G, H)')+1-g'\\&={m\over t}\cdot \ell(\gcd(G, H))-{1\over 2}\deg\textup{Diff}(F'/F)\\&={m\over t}\cdot h(\mathcal{C})-{1\over 2}\deg\textup{Diff}(F'/F).
\end{align*}
If 
	\[\deg(\gcd(G, H))>2g-2+{t\over m}\deg\textup{Diff}(F'/F),\]
then,
\begin{align*}
	\deg(\gcd(G, H)')&={m\over t}\cdot\deg(\gcd(G, H))\\&>{m\over t}(2g-2)+\deg\textup{Diff}(F'/F)\\&=2g'-2,
\end{align*}
which implies that $\gcd(G, H)'$ is non-special. So, the assertion holds.

\begin{cor}
Let $F'/\mathbb{F}_{q^t}$ be a finite Galois tame extension of an algebraic function field $F/\mathbb{F}_q$ of degree $m$. Let $P_1,\,P_2,\,\ldots,\,P_n$ be all the ramified places in $F'/F$ and $m_{P_i},\;1\le i\le n$ be the number of places of $F'$ lying over $P_i$. For an algebraic geometry code $\mathcal{C}:=C_\mathscr{L}(D, G)$ defined over $F$ with $2g-2<\deg G<n$, assume that $H$ is the divisor of $F$ such that 
\[C_\mathscr{L}(D, G)^\perp=C_\mathscr{L}(D, H).\]
 Let the divisor $\gcd(G, H)$ be non-special. If the extension $F'/F$ and the code $\mathcal{C}$ satisfies (3) and (5), then the hull of $\mathcal{C'}:=\textup{Con}_{F'/F}(\mathcal{C})$ satisfies,
\[h(\mathcal{C'})\ge{m \over t}\cdot h(\mathcal{C})-{1\over 2t}\sum\limits_{i=1}^n(m-m_{P_i}f_{P_i})\cdot \deg P_i,\]
and if
	\[\deg(\gcd(G, H))>2g-2+\sum\limits_{i=1}^n(1-{m_{P_i}f_{P_i}\over m})\cdot\deg P_i,\]
then,
		\[h(\mathcal{C'})={m \over t}\cdot h(\mathcal{C})-{1\over 2t}\sum\limits_{i=1}^n(m-m_{P_i}f_{P_i})\cdot \deg P_i.\]
 where $f_{P_i}$ is the relative degree of $P_i$.
\end{cor}

\proof By Theorem 3, 
\begin{align*}
\deg(\text{Diff}(F'/F)) &= \deg\left(\sum\limits_{i=1}^n\sum\limits_{P'|P_i}(e(P_i)-1)\cdot\deg P'\right)=\sum\limits_{i=1}^n\left (\deg(\textup{Con}_{F'/F}(P_i))-\sum\limits_{P'|P_i}\deg P'\right)\\ &=\sum\limits_{i=1}^n\left({m \over t}\deg P_i - {1\over t}\sum_{P'|P_i}f(P_i)\deg P_i\right)={1\over t}\sum\limits_{i=1}^n(m-m_{P_i}f_{P_i})\cdot \deg P_i.
\end{align*}
So, the statements follows immediately from Theorem 4.1.

\begin{cor}
Use the same notations and condition of Corollary 4.2. Let $R$ be the divisor defined as $R=P_1+\cdots P_n$. If all $P_1,\,P_2,\,\ldots,\,P_n$ are totally ramified, then the dimension of the hull of $\mathcal{C'}:=\textup{Con}_{F'/F}(\mathcal{C})$ satisfies,
		\[h(\mathcal{C'})\ge{m \over t}\cdot h(\mathcal{C})-{(m-1)\over 2t}\cdot\deg R,\]
and if
	\[\deg(\gcd(G, H))>2g-2+{(m-1)\over m}\cdot\deg R,\]
then,
		\[h(\mathcal{C'})={m \over t}\cdot h(\mathcal{C})-{(m-1)\over 2t}\cdot\deg R.\]
\end{cor}

\proof Since all $P_1,\,P_2,\,\ldots,\,P_n$ are totally ramified, $r_i=f_{P_i}=1$ for all $1\le i \le n$. So,
\[\sum\limits_{i=1}^n(1-{m_{P_i}f_{P_i}\over m})\cdot\deg P_i	=(1-{1\over m})\cdot\deg R.\]
Then, the assertion follows from Corollary 4.2.

\section{Conorm codes of rational AG codes}
A rational AG code is an algebraic geometry code defined on a rational function field $\mathbb{F}_q(x)$. For any algebraic function field $F$ over $\mathbb{F}_q$, as $F$ is an algebraic extension of $\mathbb{F}_q(x)$, the conorm code of a rational AG code is regarded as an AG code over $F$. We determine the hull dimensions of the conorm codes of rational AG codes on specific types of function fields.

Given a rational function field $\mathbb{F}_q(x)$, let $P_{\alpha}$ be the zero of $x-\alpha$ for $\alpha\in \mathbb{F}_q(x)$ and $P_\infty$ be the infinite place of $\mathbb{F}_q(x)$.
For an integer $n\ge 2$ that divides $q-1$, let $\zeta\in\mathbb{F}_q$ be a primitive $n$-th root of unity. We consider a two-point rational AG code defined as
$\mathcal{C}_{ab}=\mathcal{C}_\mathscr{L}(D, aP_0+bP_\infty)$,	
where $D=P_1+P_\zeta+\cdots +P_{\zeta^{n-1}}$ and $a,\,b\in\mathbb{Z}$ such that $0\le a+b\le n-2$ and $0\le b-a\le n$. The dual of $\mathcal{C}_{ab}$ is given by
\[\mathcal{C}_{ab}^\perp=C_\mathscr{L}(D, -(a+1)P_0+(n-b-1)P_\infty)\]
which is proved in \cite{Stichtenoth}.
\begin{prop}
The dimension of the hull of $\mathcal{C}_{ab}$ is given by
\[h(\mathcal{C}_{ab})=\begin{cases}
n-a-b-1, & \mbox{if }a\ge 0\mbox{ and } b\ge {n-1\over 2}\\
b-a, & \mbox{if }a\ge 0\mbox{ and } b< {n-1\over 2}\\
n+a-b, & \mbox{if }a< 0\mbox{ and } b\ge {n-1\over 2}\\
a+b+1, & \mbox{if }a< 0\mbox{ and } b< {n-1\over 2}.\\
\end{cases}\]
\end{prop}
\proof Assume $a\ge 0$ and $b\ge {n-1\over 2}$. Let $G=aP_0+bP_\infty$ and $H=-(a+1)P_0+(n-b-1)P_\infty$. Since $G\ge H$, $A:=\gcd(G, H)=H$. So
\begin{align*}
	\deg(A)=\deg(H)&=n-a-b-2\\&\ge 0>-2=g_0-2,
\end{align*}
where $g_0=0$ is the genus of $\mathbb{F}_q(x)$. Since this implies that $A$ is non-special,
\begin{align*}
h(\mathcal{C}_{ab})&=\deg(A)+1-g_0\\&=n-a-b-1.	
\end{align*}
For the remaining three cases, we can easily check that the divisor $\gcd(G, H)$ is non-special. So $h(\mathcal{C}_{ab})$ is immediately given by (1).

For the following three examples, we assume that (3) and (5) always holds for the algebraic function field $F$ given in examples and the code $\mathcal{C}_{ab}$ defined above.

\begin{example} [Elliptic function field]
	{\em 
	Let $F$ be an elliptic function field over $\mathbb{F}_q$. If $1<b-a<n-1$, and $0<a+b<n-2$ then, we show
\[h(\textup{Con}_{F/\mathbb{F}_4(x)}(\mathcal{C}_{ab}))=2h(\mathcal{C}_{ab})-2.\]}
\end{example}
By Proposition 6.1.2 of \cite{Stichtenoth}, there exists an equation of the form $y^2=f(x)$ or $y^2+y=f(x)$ where $f(x)\in \mathbb{F}_q(x)$, such that $F=\mathbb{F}_q(x, y)$. So $[F:\mathbb{F}_q(x)]=2$. Since 
	\[\deg \textup{Con}_{F/\mathbb{F}_q(x)}(A)=2\deg A,\]
it is sufficient to show that the degree of $A$ is positive to verify that $\textup{Con}_{F/\mathbb{F}_q(x)}(A)$ is non-special.

Suppose that  $a\ge 0$ and $b\ge {n-1\over 2}$. Since $a+b<n-2$ by assumption, $\deg A >0$. For other three cases of $a$ and $b$, we can simply check that the degree of $A$ is positive. Therefore,
\begin{align*}
	h(\textup{Con}_{F/\mathbb{F}_q(x)}(\mathcal{C}_{ab}))&=\ell(\textup{Con}_{F/\mathbb{F}_q(x)}(A))\\&=\deg \textup{Con}_{F/\mathbb{F}_q(x)}(A)+1-g\\&=2\deg A\\&=2(h(\mathcal{C}_{ab})-1).
\end{align*}

\begin{example} [Hyperelliptic function field] {\em
	Let $F=\mathbb{F}_q(x, y)$ be a hyperelliptic function field defined by an equation
	\[y^2+h(x)y=f(x),\]
where $f(x)\in\mathbb{F}_q[x]$ is a square-free polynomial of degree $m > 4$ and $\deg h=\left\lceil{m\over 2}\right\rceil+1$. If $a$ and $b$ satisfy
\[\begin{cases}
n-\left\lceil{m\over 2}\right\rceil>a+b, & \mbox{if }a\ge 0 \mbox{ and }b\ge{n-1\over 2},\\
b-a>\left\lceil{m\over 2}\right\rceil-1, & \mbox{if }a\ge 0 \mbox{ and }b<{n-1\over 2},\\
n-\left\lceil{m\over 2}\right\rceil+1>b-a, & \mbox{if }a< 0 \mbox{ and }b\ge{n-1\over 2},\\
a+b>\left\lceil{m\over 2}\right\rceil-2, & \mbox{if }a< 0 \mbox{ and }b<{n-1\over 2},\\
\end{cases}\]
then, we show
\[h(\textup{Con}_{F/\mathbb{F}_q(x)}(\mathcal{C}_{ab}))=2h(\mathcal{C}_{ab})-\left\lceil{m\over 2}\right\rceil.\]}
\end{example}
By the definition of hyperelliptic curve, genus of $F$ is given by $m=2g+1$ or $m=2g+2$,	
that is, $g=\lceil{m\over 2}\rceil-1$. If $a$ and $b$ satisfy the first condition, that is, $n-\lceil{m\over 2}\rceil>a+b$, then 
\[\deg \textup{Con}_{F/\mathbb{F}_q(x)}(A)=2\deg A> 2\lceil{m\over 2}\rceil-4=2g-2.\]
For the remaining three conditions, we can check similarly. Since $\textup{Con}_{F/\mathbb{F}_q(x)}(A)$ is non-special,
\begin{align*}
	h(\textup{Con}_{F/\mathbb{F}_q(x)}(\mathcal{C}_{ab}))=2h(\mathcal{C}_{ab})-\left\lceil{m\over 2}\right\rceil.
\end{align*}

\begin{example} [Hermitian function field] {\em
	Let $q$ be a square number, that is, $q=p^2$ for some integer $p$. Let $F=\mathbb{F}_q(x, y)$ be a Hermitian function field defined by the equation
	\[y^p+y=x^{p+1}.\]
If $a$ and $b$ satisfy
\[\begin{cases}
n-p+{2\over p}-1>a+b, & \mbox{if }a\ge 0 \mbox{ and }b\ge{n-1\over 2},\\
b-a>p-{2\over p}, & \mbox{if }a\ge 0 \mbox{ and }b<{n-1\over 2},\\
n-p+{2\over p}>b-a, & \mbox{if }a< 0 \mbox{ and }b\ge{n-1\over 2},\\
a+b>p-{2\over p}-1, & \mbox{if }a< 0 \mbox{ and }b<{n-1\over 2},\\
\end{cases}\]
then, we show
\[h(\textup{Con}_{F/\mathbb{F}_q(x)}(\mathcal{C}_{ab}))=ph(\mathcal{C}_{ab})-{p(p+1)\over 2}+1.\]}
\end{example}
\proof Since $F$ is a hermitian function field, the genus of $F$ is given by $g={p(p-1)\over 2}$, and $[F:\mathbb{F}_q(x)]=p$. For $a$ and $b$ which satisfying the first condition, we have
\begin{equation}
	\deg \textup{Con}_{F/\mathbb{F}_q(x)}(A)=p\deg A> p(p-1)-2=2g-2.
\end{equation}
Obviously, (6) holds for $a$ and $b$ satisfying any of the remaining three conditions. Since $\textup{Con}_{F/\mathbb{F}_q(x)}(A)$ is non-special,
\begin{align*}
	h(\textup{Con}_{F/\mathbb{F}_q(x)}(\mathcal{C}_{ab}))=ph(\mathcal{C}_{ab})-{p(p+1)\over 2}+1.
\end{align*}

\section{Conclusion}
In this paper, we have determined the hull dimensions of conorm codes of algebraic geometry codes when the extension $F'/\mathbb{F}_{q^t}$ over $F/\mathbb{F}_q$ is separable, either unramified or ramified. Consequently, we have obtained the exact hull dimensions of the conorm codes of algebraic geometry codes $\mathcal{C} = C_\mathscr{L}(D, G)=C_\mathscr{L}(D, H)^\perp$ in the case where $\deg \gcd(G, H)>2g-2+{t\over [F':F]}\deg\text{Diff}(F'/F)$. Considering unramified extensions, we have showed that the conorm of an LCD code $\mathcal{C} = C_\mathscr{L}(D, G)$ such that $\gcd(G, H)$ is non-special is also an LCD code if the code $\mathcal{C}$ is defined over an elliptic function field. For an AG code defined over a non-elliptic function field that is LCD, studying the hull of its conorm may be the subject for future work of this paper.

\end{document}